\documentstyle[epsf]{mn}
\newif\ifAMStwofonts
\AMStwofontstrue

\title[Helium line enhancement by non-thermal motions]{Enhancement of
the helium resonance lines in the solar atmosphere by suprathermal
electron excitation I: non-thermal transport of helium ions} 
\author[Smith \& Jordan]{G.R.~Smith\footnotemark, C.~Jordan \\ 
Department of Physics (Theoretical Physics), Oxford University, 1 Keble Road,
Oxford OX1 3NP, UK}
\date{8 August 2002}
\pagerange{}
\pubyear{}

\begin{document}

\maketitle

\label{firstpage}

\begin{abstract}
Models of the solar transition region made from lines other than those
of helium cannot account for the strength of the helium lines. 
However, the collisional excitation rates of the helium resonance lines are
unusually sensitive to the energy of the exciting electrons. 
Non-thermal motions in the transition region could drive
slowly-ionizing helium ions rapidly through the steep
temperature gradient, exposing them to excitation by electrons
characteristic of higher temperatures than those describing their
ionization state. We present the results of
calculations which use a more physical representation of the
lifetimes of the ground states of He~{\sc i} and He~{\sc ii} than was
adopted in earlier work on this process. New emission measure
distributions are used to calculate the temperature variation with
height. The results show that non-thermal motions can lead to
enhancements of the He~{\sc i} and He~{\sc ii} resonance line
intensities by factors that are comparable with those 
required. Excitation by non-Maxwellian electron distributions would
\emph{reduce} the effects of non-thermal transport. 
The effects of non-thermal motions are more consistent with
the observed spatial distribution of helium emission than are those of
excitation by non-Maxwellian electron distributions alone. In particular,
they account better for the observed line intensity ratio
$I$(537.0~\AA)/$I$(584.3~\AA), and its variation with location.
\end{abstract}

\begin{keywords}
line: formation -- Sun: atmospheric motions -- Sun: transition region
-- Sun: UV radiation.
\end{keywords}

\section{Introduction}
\footnotetext{E-mail: g.smith2@physics.ox.ac.uk}
The resonance lines of He~{\sc i} (at 584.3 \AA) and He~{\sc ii} (at 303.8
\AA) show unusual behaviour when compared with other strong emission
lines in solar {\sc euv} spectra. Reviews of observations of the
helium resonance lines and attempts to model their formation can be
found in Hammer \shortcite{rh97} and Macpherson \& Jordan
\shortcite{mj99} (hereafter MJ99), but the details relevant to
the present work are summarized here. 

Early observations (e.g.\ Tousey 1967) showed that the helium
resonance lines are reduced significantly in intensity in coronal
holes compared with the average quiet Sun, while other lines formed
at similar temperatures show only very small reductions in intensity. 
In the quiet Sun, the helium resonance line intensities are too large to be
reproduced by emission measure distributions that account for other
transition region (TR) lines, a problem identified by Jordan (1975), who
found disagreements by factors of 15 for He~{\sc i} and 6 for He~{\sc
ii}. Corresponding factors of at least 10 and 13, respectively, were
found by MJ99.
Radiative transfer calculations have also been unable to
reproduce the observed resonance line intensities without invoking
either a `plateau' in the model atmosphere (see e.g. Vernazza, Avrett
\& Loeser 1981; Andretta \& Jones 1997), which leads to other lines
formed in the lower TR being over-estimated \cite{grs00}, or some
departure from equilibrium (e.g.\ Fontenla, Avrett \& Loeser 1993;
Avrett 1999). These results have been taken to imply that 
some process preferentially enhances the helium resonance line
intensities with respect to other lines formed in the TR in the quiet
Sun, and that the enhancement is reduced in coronal holes. Recent
results from the Solar and Heliospheric Observatory ({\em SOHO}) show
that the helium resonance line intensities are reduced by factors of
1.5--2.0 (Peter 1999; Jordan, Macpherson \& Smith 2001) in coronal holes.

Many possible enhancement processes have been investigated, but a
completely convincing explanation has not yet been
found. The effects of photoionization by coronal radiation have been
suggested as a natural explanation of the coronal hole/quiet Sun
contrast \cite{hz75}, and photoionization-recombination (PR) appears to
be important in 
the formation of some lines in the helium spectrum (He~{\sc ii} 1640.4
\AA\ -- Wahlstr\o m \& Carlsson 1994, He~{\sc i} 10830 \AA\ --
Andretta \& Jones 1997). Although PR probably contributes to the
formation of the He~{\sc i} 584.3-\AA\ line, evidence \cite{rm75,aj,and99}
suggests that the He~{\sc ii} 303.8-\AA\ line and, to a lesser extent,
the 584.3-\AA\ line, are formed principally by collisional excitation. 
If this is the case, then because the helium resonance lines have
unusually large values of $W/kT_{\textrm{e}}$, where $W$ is the
excitation energy, 
their collisional contribution functions are sensitive to excitation
by suprathermal electrons. Any process exposing helium ions to larger
populations of suprathermal electrons than in equilibrium will tend to
increase the collisional excitation rates of the helium lines, while
lines with smaller $W/kT_{\textrm{e}}$ will be relatively unaffected. 

Two processes by which this might occur were suggested by Jordan
(1975,1980): the transport of high energy electrons from the
upper TR or corona to the lower TR, or the transport of He atoms and ions
by turbulent motions to regions of higher electron temperature. Both
processes would be expected to depend on the magnitude of the
temperature gradient, which could explain the coronal hole/quiet Sun
contrast, since Munro \& Withbroe \shortcite{mw72} found d$T/$d$h$ to
be an order of magnitude smaller inside coronal holes.

Shoub \shortcite{es83} investigated the former process in his study
of the shape of the electron velocity distribution function (EVDF) in
the transition region. His calculations suggested that EVDFs in the
lower TR should have more heavily populated suprathermal tails than
the local Maxwellian distribution. He found that this could lead to
enhanced collisional excitation and ionization in helium, but did
not calculate intensities to compare with observations. Anderson,
Raymond \& Ballegooijen (1996) did calculate intensities for the He~{\sc ii}
resonance line under excitation by several different non-Maxwellian
EVDFs, and found enhancements over Maxwellian collisional excitation,
but their calculations could not reproduce the observed intensities of
all the lines they studied simultaneously. 

Radiative transfer calculations of the helium resonance line
intensities using EVDFs approximating those derived by Shoub
(1982,1983) have been performed, and are reported in a companion
paper (Smith, in preparation -- hereafter paper II). The calculations
suggest that the He~{\sc i} and He~{\sc ii} resonance line intensities
could be increased by non-Maxwellian collisional excitation, but that
this process would produce signatures in the line ratios that
contradict observations.  

The second process is the subject of
this paper. The temperature gradients in the solar TR are such that
velocities of order 10 km s$^{-1}$ could carry material into regions
of significantly higher electron temperature. Such mixing could be 
caused by non-thermal motions in the TR that can be inferred
from observations of the excess widths of optically thin
lines (e.g.\ Berger, Bruner \& Stevens 1970; Boland et al.\ 1973;
Doschek et al.\ 1976; 
Chae, Sch\"uhle \& Lemaire 1998. See also Section
\ref{sec2.3}). Helium ions would reach 
statistical equilibrium at the new local temperature much more
slowly than the bulk of the material owing to their long excitation
and ionization times.  
The helium resonance lines would be excited collisionally at
higher temperatures and hence with greater rates than would be the
case in statistical equilibrium at the temperatures determining the
ground state populations, while other transition region lines are
relatively unaffected. Jordan \shortcite{cj80} investigated the effects of
the process on the He~{\sc ii} 303.8-\AA\ line, and a more extensive
study has been made recently by Andretta et al.\ \shortcite{aea00}, who
termed the process `velocity redistribution.' 

This work represents the application of similar methods to more
specific examples of temperature gradients, which are no longer
assumed to be constant over the path of the moving clump of plasma. A
new treatment is used for the mean lifetime of the helium ground state,
and the analysis is extended to the cases of the He~{\sc i} 584.3-\AA\
and 537.0-\AA\ lines in a first approximation.  

In Section \ref{sec2} earlier work on non-thermal transport of helium is
reviewed, and the methods and atmospheric parameters used here to
calculate the enhancements of the helium line intensities are
introduced. In Section \ref{sec3} the calculations of the velocity
redistribution enhancement factors for the \hbox{He\,{\sc ii}}
303.8-\AA\ line and the \hbox{He\,{\sc i}} 584.3-\AA\ and 537.0-\AA\ lines
are described in detail. The possible effects of non-Maxwellian
electron distributions in the transition region on the calculations
are assessed. In Section \ref{sec4} the results of the calculations
are discussed. The derived enhancement factors are 
compared with the results of MJ99, to determine whether the process
can explain quantitatively the anomalously large helium
line intensities. More qualitative comparisons are made with
observations of the spatial variations of the helium line intensities
in the quiet Sun with respect to each other and to other transition
region lines. Comparisons are
also made with the results of a separate investigation into the
possibility that the helium resonance line intensities are enhanced by
excitation by non-local suprathermal electrons (Paper II).

\section{Methods of calculation}
\label{sec2}
\subsection{Earlier work}
Jordan (1980) investigated the enhancement that would occur in
the He~{\sc ii} 303.8-\AA\ line intensity if the line were collisionally
excited at a higher temperature than that appropriate to conditions of 
statistical equilibrium. Starting at an initial temperature of
$T_{\textrm{i}} = 8 \times 10^{4}$ K, she assumed that He~{\sc ii}
ions are carried 
up the temperature gradient at the local non-thermal (macroturbulent)
velocity derived from the excess widths observed in lines formed near
the initial temperature. She wrote the
excitation time of the ion's ground state as the reciprocal of the
collisional excitation rate (at the final temperature -- see below for
a discussion of this point). The excitation time and the non-thermal
velocity were used to calculate a free path travelled by the ion
before it is excited. The final temperature of excitation was
then computed using a mean temperature gradient. Using
values of the non-thermal velocity and the temperature gradient 
appropriate to different regions of the atmosphere (quiet Sun, coronal
holes, active regions), she found intensity
enhancements of up to a factor of 5. This was of the right
order needed to explain the discrepancy between observed and
calculated intensities reported in Jordan \shortcite{cj75}.

Andretta et al.\ \shortcite{aea00} followed the same approach for the
He~{\sc ii} resonance line, but
instead of a using single upward non-thermal velocity, they considered an
ensemble of plasma elements (clumps) in an isotropic turbulent
velocity field with a Gaussian distribution having a standard deviation
equal to the root-mean-squared non-thermal velocity $\sqrt{\langle
v_{\textrm{\sc t}}^{2} \rangle}$. Andretta et al.\ (2000) also used
$T_{\textrm{i}} = 8 
\times 10^{4}$ K and assumed a constant pressure in the transition
region. They calculated the enhancement factor over a
parameter space of $\langle v_{\textrm{\sc t}}^{2} \rangle$ and d$T/$d$h$
appropriate 
to conditions in the TR, and for values of the electron pressure
corresponding to quiet Sun and active region conditions. 
Enhancement factors of up to about 5 were obtained for values of the
parameters typical of current models and observations of the quiet Sun
transition region ($P_{\textrm{e}}/k = 5 \times 10^{14}$ cm$^{-3}$ K,
$\sqrt{\langle v_{\textrm{\sc t}}^{2} \rangle} \geq 10$ km s$^{-1}$,
d$T/$d$h \geq 10^{3}$ K km$^{-1}$). Scaling arguments derived from their
analysis show reasonable agreement with observations of the
\hbox{He\,{\sc ii}} 303.8-\AA\ line compared with the \hbox{O\,{\sc
iii}} 599.6-\AA\ line. They concluded that the 
process could account for at least some of the enhancement
required in the 303.8-\AA\ line intensity.

Our calculations of the effects of non-thermal motions
follow the approaches of Jordan \shortcite{cj80} and Andretta et
al. \shortcite{aea00}, but with certain changes. The above studies used
the `coronal approximation,' assuming that in the resonance lines
radiative decay is balanced by collisional excitation. Radiative
transfer calculations suggest that this is certainly realistic in the
case of the He~{\sc ii} resonance line \cite{grs00}, so this
assumption is also made here. In treating the He~{\sc i} lines, 
advantage is taken of information from radiative transfer calculations
made using full atmospheric models.

In the coronal approximation the enhancement factor in the intensity
of the 303.8-\AA\ line is given approximately by the ratio of the
collisional excitation rate at the final temperature to that at the
initial temperature. The collisional excitation rate $C_{ij}$ is
given by the expression
\begin{equation}
\label{eq0}
C_{ij}(T_{\textrm{e}}) = 8.63 \times 10^{-6} \frac{\Omega_{ij}}{g_{i}}
T_{\textrm{e}}^{-1/2} N_{\textrm{e}} \textrm{ exp} \left
( -\frac{W}{kT_{\textrm{e}}} \right) \textrm{ s}^{-1} 
\end{equation}
where $g_{i}$ is the statistical weight of the lower level
and $\Omega_{ij}$ is the collision strength, which is approximately
constant over the temperature range of interest. The intensity
enhancement ratio is therefore given approximately by 
\begin{equation}
\label{eq1}
\frac{I_{\textrm{f}}}{I_{\textrm{i}}} \simeq
\frac{C_{ij}(T_{\textrm{f}})}{C_{ij}(T_{\textrm{i}})}. 
\end{equation}
Assuming the electron pressure to be constant in the
transition region, using equation (\ref{eq0}), this becomes 
\begin{equation}
\label{eq2}
\frac{Cij_(T_{\textrm{f}})}{C_{ij}(T_{\textrm{i}})} = \left
( \frac{T_{\textrm{i}}}{T_{\textrm{f}}} \right)^{3/2} 
\textrm{exp} \left( \frac{W}{k} \left( \frac{1}{T_{\textrm{i}}} -
\frac{1}{T_{\textrm{f}}} \right) \right).
\end{equation}

The final temperature $T_{\textrm{f}}$ depends on the velocity of the
emitting plasma in the direction of the temperature gradient. In order to
generalize the calculation of the enhancement factor to an ensemble of
plasma elements Andretta et al.\ (2000) convolved equation (\ref{eq2})
with a normalized Gaussian distribution of turbulent velocities with
standard deviation $\sqrt{\langle v_{\textrm{\sc t}}^{2} \rangle}$:
\begin{equation}
G(v) = \frac{1}{\sqrt{2\pi\langle v_{\textrm{\sc t}}^{2} \rangle}}
\textrm{ exp} \left( -\frac{1}{2} \frac{v^{2}}{\langle v_{\textrm{\sc
t}}^{2} \rangle} \right)
\end{equation} 
to give an intensity ratio of
\begin{equation}
\label{eq3}
\frac{I_{\textrm{f}}}{I_{\textrm{i}}} = \int_{-\infty}^{\infty}
\frac{C_{ij}(T_{\textrm{f}})}{C_{ij}(T_{\textrm{i}})}G(v)\textrm{d}v.
\end{equation} 

\subsection{New formulations} 
Equation (\ref{eq3}) was used in calculations of the enhancement
factors in the present work, but the free path and final temperature
$T_{\textrm{f}}$ of each plasma element were computed by different methods
to those used by Jordan (1980) and Andretta et al.\ (2000). These new
techniques use
the temperature gradient (now allowed to change over
the path of the clump of plasma) derived from emission measure
distributions found from observations of other transition region
lines. Specific values of $\langle v_{\textrm{\sc t}}^{2}
\rangle(T_{\textrm{i}})$ suggested by observations were
used. For the 584.3-\AA\ and 537.0-\AA\ lines of He~{\sc i} the
excitation rate ratio in equation (\ref{eq2}) was also modified. 

Both Jordan (1980) and Andretta et al.\ (2000) defined the lifetime of
the ground state as $\tau = (C_{ij}(T_{\textrm{f}}))^{-1}$. This
provides a lower limit on the lifetime and hence on the final temperature
reached by the moving plasma. This may be appropriate if ions are
carried rapidly to a higher temperature and remain there until
excited. Here we use $\tau =
(C_{ij}(T_{\textrm{i}}))^{-1}$, as we consider this timescale to
better reflect the dependence of the enhancement process in question
on the long excitation and ionization times of the helium ions at the
equilibrium temperatures of resonance line formation. It is these
timescales, long with respect to those for other TR lines, which allow
unresolved turbulent motions to carry helium ions to the higher
temperatures where excitation occurs. 

$T_{\textrm{i}}$ is chosen to represent the `normal' temperature of
resonance line formation, in the absence of non-equilibrium processes.
$T_{\textrm{f}}$ is found by integrating the temperature gradient over
the path traversed in time $\tau$. The (varying) temperature gradient
between $T_{\textrm{i}}$ and $T_{\textrm{f}}$ is 
derived from the mean intrinsic emission measure distribution (EMD;
see Section \ref{sec2.3}). The intrinsic 
emission measure is defined as 
\begin{equation}
Em(0.3) = \int_{\Delta h'} N_{\textrm{e}}N_{\textrm{\sc h}}\textrm{d}h
\end{equation}
where $N_{\textrm{e}}$ and $N_{\textrm{\sc h}}$ are electron and hydrogen
number densities respectively and $\Delta h'$ is defined as the
interval of height in a plane-parallel atmosphere corresponding to
$\Delta$(log~$T_{\textrm{e}}$) = 0.3 dex, the temperature interval
over which most lines are typically formed. Defining electron and
hydrogen pressures as $P_{\textrm{e}} = N_{\textrm{e}}T_{\textrm{e}}$
and $P_{\textrm{\sc h}} = N_{\textrm{\sc h}}T_{\textrm{e}}$, the
emission measure may be re-written as an integral over the temperature
interval in which a line is formed:
\begin{equation}
Em(0.3) = \int_{T'_{e}/\sqrt{2}}^{\sqrt{2}T'_{e}}
\frac{P_{\textrm{e}}P_{\textrm{\sc h}}}{T_{\textrm{e}}^{2}}
\frac{\textrm{d}h}{\textrm{d}T_{\textrm{e}}} \textrm{d}T_{\textrm{e}}.
\end{equation}
$T'_{\textrm{e}}$ is the temperature at which the line contribution
function peaks, and is the geometric mean of the limits of the
integral, which represent a change in $T_{\textrm{e}}$ of a factor of
2. If $P_{\textrm{e}}$, $P_{\textrm{\sc h}}$, and
$\textrm{d}h/\textrm{d}T_{\textrm{e}}$ are each averaged over the
temperature range of the integral, the emission measure becomes
\begin{equation}
\label{eq4}
Em(0.3) =
\frac{\overline{P_{\textrm{e}}} \overline{P_{\textrm{\sc
h}}}}{\sqrt{2}T_{\textrm{e}}}
\overline{\frac{\textrm{d}h}{\textrm{d}T_{\textrm{e}}}}. 
\end{equation}
The assumption of constant pressures over the region of line formation
is a good one, and although the assumption of a a constant temperature
gradient in the same region is less good, it is often used in
constructing an initial model from the EMD (see e.g.\ Jordan \& Brown
1981). 

If the mean EMD as a function of temperature is known from several
lines, the path of material travelling up the temperature gradient can be
divided into smaller temperature intervals. Here we use intervals
of 0.1 dex in log~$T_{\textrm{e}}$. In each of
these intervals, the mean temperature gradient is calculated from the
EMD, at the mean value of log~$T_{\textrm{e}}$, using equation
(\ref{eq4}). In most of the temperature range of interest, hydrogen is
fully ionized and $N_{\textrm{\sc h}} = 0.83 N_{\textrm{e}}$. This is a
very good approximation above log~$T_{\textrm{e}}$ = 4.9 (i.e. in the
He~{\sc ii} calculation), and is correct to better than 10 per cent
down to log~$T_{\textrm{e}}$ = 4.5 (the 
temperature range of the He~{\sc i} calculation). Thus, in
plane-parallel geometry, taking $P_{\textrm{\sc h}} = 0.83
P_{\textrm{e}}$, the increment of height $\Delta h$ corresponding to
the temperature interval $\Delta$(log $T_{\textrm{e}}$) is given
approximately by  
\begin{equation}
\label{eq6}
\Delta h = Em(0.3) \frac{\sqrt{2}T_{\textrm{e}}}{0.83
P_{\textrm{e}}^{2}} \Delta T_{\textrm{e}} 
\end{equation}
where $\Delta T_{\textrm{e}}$ spans the range log~$T_{\textrm{e}} \pm
0.05$. The 
travel time required by a clump of material to traverse each $\Delta h$
is determined by assuming a constant upward velocity $v$. 
The definition of $\Delta h$ thus introduces a quantization of
the (logarithmic) temperatures used in calculating the
enhancement factor. For computational convenience this is carried
over into the integration over velocities in equation
(\ref{eq3}). In this formulation, the integrand may only be evaluated
correctly for values of $T_{\textrm{f}}$ occurring at intervals of 0.1 dex in
log~$T_{\textrm{e}}$ from the initial temperature $T_{\textrm{i}}$
(log~$T_{\textrm{f}}$ = log~$T_{\textrm{i}}$ + 0.1$n$, for integer
$n$). The values of $v$ used as points
in the integration correspond to velocities required to reach these
$T_{\textrm{f}}$ in a time equal to the lifetime of the ion's ground
state at $T_{\textrm{i}}$. The velocity mesh that results introduces
some uncertainty into the calculation of the enhancement factors. 

\subsection{Atmospheric parameters}
\label{sec2.3}
The EMDs used here to derive the variation of temperature with height
in the lower TR
were determined from observed line fluxes and the appropriate atomic
data using now-standard methods (see MJ99 and references therein).
MJ99 derived separate EMDs for mean network and cell interior regions
from observations of UV lines from the quiet Sun transition region
made using the {\sc cds} (Coronal Diagnostic Spectrometer) and {\sc
sumer} (Solar Ultraviolet
Measurements of Emitted Radiation) instruments on board {\em SOHO}.
Equation (\ref{eq4}), used to determine the temperature gradient
from the emission measure, assumes a one-component plane-parallel
atmosphere. In the network model proposed by Gabriel \shortcite{ag76},
open magnetic field lines are most vertical over the boundaries, but
expand into the corona with a horizontal component in the cell
interiors. Since we adopt a plane parallel atmosphere we use the
network boundary EMD of MJ99 as the observed EMD.

Of the lines observed by MJ99, those formed at temperatures below
$10^{5}$ K (except for He~{\sc ii}) were observed with {\sc sumer},
and those formed at higher temperatures (and the helium lines) were
observed with {\sc cds}. MJ99 found
that the relative intensities of the He~{\sc i} lines observed with
{\sc cds} and with {\sc sumer} in second order suggest that the {\sc
sumer} lines might need to be reduced in intensity by a factor of 1.5
if the Landi et al.\ \shortcite{lan97} {\sc cds} calibration were
adopted. This would lead to the emission measure distribution below
$10^{5}$ K, derived from the {\sc sumer} lines, being lowered relative
to the higher temperature part derived from the {\sc cds} lines. 

This uncertainty inspired the use of a second EMD, derived from
observations of $\xi$ Boo A, a G8 V star, made with the Goddard High
Resolution Spectrograph (GHRS) instrument on the Hubble Space
Telescope \cite{grs00}. The EMD was scaled by a constant factor such
that its minimum (at 1 -- $2 \times 10^{5}$ K) fits the minimum of the
network emission measure of MJ99. The resulting EMD is lower, by up to
an order of magnitude, at temperatures below log~$T_{\textrm{e}} =
4.8$, but is very similar to the solar EMD at higher temperatures. 
Jordan et al. \shortcite{jea87} found that below
$T_{\textrm{e}} \sim 10^{5}$ K the emission measure distributions of
main sequence stars with hot coronae have approximately the same
dependence on $T_{\textrm{e}}$, which is supported by a direct linear
correlation between the C~{\sc ii} and C~{\sc iv} emission line fluxes
in such stars \cite{rea91}. Capelli et al.\ \shortcite{cea89} found
that the C~{\sc ii} -- C~{\sc iv} correlation also holds for solar
observations at various spatial resolutions.  

\begin{figure}
\centering
\begin{minipage}{3.6in}
\epsfxsize=3.6in \epsfbox{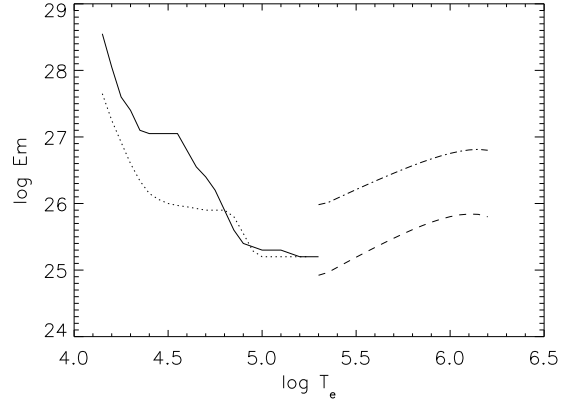}
\end{minipage}
\caption{Network emission measure distributions. For
log~$T_{\textrm{e}} \leq 5.3$, the solid line shows EMD S, the dotted
line shows EMD X. Theoretical EMDs are shown for log~$T_{\textrm{e}} >
5.3$, with (a) $P_{\textrm{e}}($log~$T_{\textrm{e}} = 5.3) = 2 \times
10^{14}$ cm$^{-3}$ K (dashed), and with (b)
$P_{\textrm{e}}($log~$T_{\textrm{e}} = 5.3) = 6 \times 10^{14}$ cm$^{-3}$
K (dot-dashed).\label{fig1}}
\end{figure}

The discrepancy between the {\em SOHO} observations and the earlier
results may be due in part to the different spatial resolutions of 
the two instruments used ({\sc cds} and {\sc sumer}).
A more serious problem is that {\sc sumer} could not observe the same
location on the Sun in all of the lower TR lines, or the same location
at the same time as in the {\sc cds} observations (see Jordan et al.\
2001). Thus there are fundamental uncertainties in the shape of the
EMD derived from {\em SOHO} observations at temperatures below
log~$T_{\textrm{e}} \simeq 4.9$. The $\xi$ 
Boo A EMD was therefore used to explore the possible effects of this
uncertainty on the present calculations. The solar EMD derived by MJ99
is hereafter referred to as EMD S, and that based on the $\xi$ Boo A
EMD is referred to as EMD X; they are shown in Figure~\ref{fig1}.

The lines observed by MJ99 did not allow determination of the full EMD
above log~$T_{\textrm{e}}$ = 5.3. The EMDs used in the present calculations
above this temperature are therefore based on energy balance arguments
described below (see also Jordan 2000 and Philippides 1996). 

In the lower TR, below log~$T_{\textrm{e}} \simeq 5.0$, some non-thermal
source of heating is required to account for the rising emission
measure at low temperatures. In the upper TR, however, the radiation
losses are small compared with the conductive flux from the corona,
and may be balanced by energy deposited by the divergence of the
conductive flux, or, if that flux is conserved, by a small amount of
additional heating. The EMDs used here were determined assuming the
former to be the case above log~$T_{\textrm{e}}$ = 5.3. The equation
expressing energy balance as a function of height in this region is then 
\begin{equation}
\label{eq5}
\epsilon_{\textrm{\sc r}} = -\frac{1}{A(r)}
\frac{\textrm{d}(A(r)F_{\textrm{\sc c}})}{\textrm{d}r} 
\end{equation}
where $\epsilon_{\textrm{\sc r}}$ is the rate of radiative energy loss
per unit volume, and $F_{\sc c}$ is the classical conductive energy
flux through an area $A(r)$ normal to the radial co-ordinate $r$. The
radiation loss term is given by 
\begin{equation}
\epsilon_{\textrm{\sc r}} =
N_{\textrm{e}}N_{\textrm{\sc h}} P_{\textrm{rad}}(T_{\textrm{e}}) 
\end{equation}
where $P_{\textrm{rad}}(T_{\textrm{e}})$ is the radiative power loss
in all lines and continua, as a function of temperature in a plasma of
known composition. $P_{\textrm{rad}}(T_{\textrm{e}})$ is
calculated here using the power law fit to the radiative power loss curve
adopted by Philippides \shortcite{dp96}: 
\begin{equation}
P_{\textrm{rad}}(T_{\textrm{e}}) = 1.25 \times 10^{-16} T_{\textrm{e}}^{-1}
\textrm{ erg cm}^{3}\textrm{ s}^{-1}.
\end{equation}
$F_{\textrm{\sc c}}$ is given by 
\begin{equation}
F_{\textrm{\sc c}}(T_{\textrm{e}}) = -\kappa T_{\textrm{e}}^{5/2}
\frac{\textrm{d}T_{\textrm{e}}}{\textrm{d}r} 
\end{equation}
where the Spitzer-H\"arm conductivity $\kappa \simeq 1.1 \times
10^{-6}$ erg cm$^{-1}$ K$^{-7/2}$ s$^{-1}$ at $T_{\textrm{e}} = 10^{6}$ K
\cite{ls56}. That value is used here, as $\kappa$ is only 30\% smaller
at $T_{\textrm{e}} = 10^{5}$ K. Energy transfer by mass motions of any
kind (including the turbulent motion assumed in the enhancement factor
calculations) is not considered in the energy balance equation used in
the upper TR.

Combining the energy balance equation (\ref{eq5}) with equation
(\ref{eq4}) for the emission measure in terms of the temperature
gradient gives an expression for the logarithmic gradient of the
emission measure with temperature \cite{cj00}:
\begin{eqnarray}
\label{eq8}
\frac{\textrm{d log }Em(0.3)}{\textrm{d log }T_{\textrm{e}}} &=& \frac{3}{2} +
2\frac{\textrm{d log }P_{\textrm{e}}}{\textrm{d log }T_{\textrm{e}}} +
\frac{\textrm{d log }A(T_{\textrm{e}})}{\textrm{d log }T_{\textrm{e}}}
\nonumber\\ 
 & & -\frac{2Em(0.3)^{2}P_{\textrm{rad}}(T_{\textrm{e}})}
{0.83\kappa P_{\textrm{e}}^{2}T_{\textrm{e}}^{3/2}}. 
\end{eqnarray}

Assuming forms for $P_{\textrm{e}}(T_{\textrm{e}})$ and
$A(T_{\textrm{e}})$, this equation may be integrated numerically from
given boundary conditions to give the emission measure as a function
of temperature in that part of the atmosphere where the energy balance
is described by equation (\ref{eq5}). In this case the electron
pressure was assumed to vary according to hydrostatic equilibrium
(the use of a constant value of $P_{\textrm{e}}$ in equation
(\ref{eq2}) in calculating the intensity enhancement is a good
approximation in the lower TR, owing to its small vertical extent).
$A(T_{\textrm{e}})$ was taken only to vary with the
expansion with height of a spherically symmetric atmosphere; assuming
that emission comes from a uniform layer at each height, this is
consistent with the use of the empirical EMDs derived in the
plane-parallel approximation in the lower TR, because the small extent
of that part of the atmosphere makes the two approaches approximately
equivalent. 

The assumptions about $A(T_{\textrm{e}})$ represent a significant
simplification. The EMD derived below log~$T_{\textrm{e}} = 5.3$
represents an average over any unresolved structure within the network
boundaries, while the EMD computed using equation (\ref{eq8}) is the
`intrinsic' value; they are unlikely to match at log~$T_{\textrm{e}} =
5.3$. If the calculated emission measure at log~$T_{\textrm{e}} = 5.3$
is greater than the value derived from spatially averaged
observations, this implies that only some fraction of the observed
region is occupied by emitting material.  

With the assumed forms for $P_{\textrm{e}}(T_{\textrm{e}})$ and
$A(T_{\textrm{e}})$, the EMD for
the upper TR was determined using equation (\ref{eq8}), choosing
values of the emission measure, electron pressure, and coronal
temperature $T_{\textrm{\sc c}}$ at the top of the
region. $T_{\textrm{\sc c}}$ was chosen such that log~$T_{\textrm{\sc
c}} = 6.20$, as {\sc cds} observations suggest that there 
is little material at higher temperatures in the quiet corona \cite{lal98}. 
For a given coronal pressure and temperature, the coronal emission
measure cannot exceed a certain value if the RHS of equation
(\ref{eq8}) is to remain positive above log~$T_{\textrm{e}} = 5.3$, as
is required by the observed gradient of the EMD. There is hence a `critical
solution,' with the largest possible value of the coronal emission
measure which reproduces the observed minimum in the EMD at
log~$T_{\textrm{e}} \simeq 5.3$. If $Em(T_{\textrm{\sc c}})$ is larger
than this value, conduction from the corona deposits more energy in
the upper TR than can be lost by radiation.

The mean pressure in the boundary regions used in deriving EMD S
below log~$T_{\textrm{e}} = 5.3$ was found by MJ99 to be
log~$P_{\textrm{e}} = 14.27$,
with a standard deviation of $\pm 0.35$. Using a coronal pressure that
reproduces this pressure ($\simeq 2 \times 10^{14}$ cm$^{-3}$ K) at 
log~$T_{\textrm{e}} = 5.3$ in the critical solution gives EMD (a) shown in
Figure \ref{fig1}. This is lower than the observed emission measure at
log~$T_{\textrm{e}} = 5.3$, suggesting that a slightly higher pressure is
needed for a continuous EMD through this temperature. EMD (a) also
predicts substantially less Mg~{\sc ix} and Mg~{\sc x} emission than
observed by MJ99 (when the photospheric abundance of Mg is used in
both $P_{\textrm{rad}}(T_{\textrm{e}})$ and in calculating the line
intensities). There are problems with such a comparison, however, as
the intensities used by MJ99 correspond to areas defined as network
boundaries in the {\sc cds} transition region lines, and the details
of the connection between the TR network and the corona are
unknown. Open field regions would be expected to expand in the corona,
but the network boundaries may also contain closed field emitting
regions unconnected to the corona. 

Figure \ref{fig1} also shows EMD (b), a second upper TR EMD
calculated to give a higher transition region pressure of
$P_{\textrm{e}} = 6 \times 10^{14}$ cm$^{-3}$ K, which corresponds to
the largest pressures found by MJ99 (in cell interiors) and is similar
to values used in earlier models of the atmosphere (e.g.\ Vernazza et
al.\ 1981). It provides a much better match to the Mg~{\sc ix} and
Mg~{\sc x} data, although, as stated above, these do not provide a stringent 
constraint. EMD (b) predicts an intrinsic emission measure at
log~$T_{\textrm{e}} = 5.3$ that is larger than the apparent (plane-parallel)
emission measure. This discontinuity can be explained in terms of an
area filling factor below log~$T_{\textrm{e}} = 5.3$ (see Sim \& Jordan 2001).
For a filling factor $A_{\textrm{i}}/A_{\textrm{obs}}$, the intrinsic
emission measure below log~$T_{\textrm{e}} = 5.3$ is 
\begin{equation}
\label{eq15} 
Em_{\textrm{i}} = \frac{A_{\textrm{obs}}}{A_{\textrm{i}}} Em_{\textrm{app}}
\end{equation}
where $Em_{\textrm{app}}$ is shown in Figure \ref{fig1}. In the
theoretically derived upper TR the filling factor is assumed to be
1.0. Scaling the lower TR EMD to match the upper TR EMD (b) at
log~$T_{\textrm{e}} = 5.3$ suggests a filling factor in the lower TR
of 0.16 (within the area of $\sim$ 3.4 arcsec$^2$ resolved by {\sc cds}).
The continuous EMD then describes the restricted areas where the
filling factor is 1.0.
The lines used to find the lower TR EMD are not sensitive to
$P_{\textrm{e}}$, so a similar matching procedure may be considered for
any lower pressure above about $2.7 \times 10^{14}$ cm$^{-3}$ K, for
which the theoretical upper EMD would match the observed emission
measure at log~$T_{\textrm{e}} = 5.3$, with a filling factor of 1.0.
Equations (\ref{eq4}) and (\ref{eq15}) show that the temperature
gradient derived from $Em(T_{\textrm{e}})$
scales as $P_{\textrm{e}}^{2} A_{\textrm{i}}/A_{\textrm{obs}}$, but
the value of $A_{\textrm{i}}/A_{\textrm{obs}}$ (i.e.\ the factor by
which the theoretical and observed EMDs differ at log~$T_{\textrm{e}} = 5.3$)
required for a continuous EMD through log~$T_{\textrm{e}} = 5.3$ scales as
$P_{\textrm{e}}^{-2}$. If the filling factor in the lower TR is
regarded as a free parameter, the distribution of
d$T_{\textrm{e}}/$d$h$ is the same for all continuous EMDs in which
a lower TR EMD is scaled to match an upper TR EMD, and it
is independent of the assumed pressure.  

The distributions of d$T_{\textrm{e}}/$d$h$ derived in this manner for
the cases of lower TR EMDs S and X were used in the calculations of
the helium line enhancement factors. In order to investigate the
effect of the possible solutions suggested by EMDs (a) and (b), the
calculations were performed for electron pressures representing these
limits, $P_{\textrm{e}} = 2 \times 10^{14}$ cm$^{-3}$ K and
$P_{\textrm{e}} = 6 \times 10^{14}$ cm$^{-3}$
K (the lower pressure is not technically consistent with the filling
factor argument, but, considering other approximations involved in the
calculations, provides a useful lower limit). Because the temperature
gradients used in the enhancement factor calculations are independent
of the assumed pressure, the pressure adopted influences only the 
excitation time. The excitation time scales as $P_{\textrm{e}}^{-1}$;
hence the velocity required to reach a given $T_{\textrm{f}}$,
$v(T_{\textrm{f}})$, scales as $P_{\textrm{e}}$. A higher pressure
will therefore lead to larger $v(T_{\textrm{f}})$, 
causing contributions from higher temperature regions to be suppressed
by the Gaussian term, $G(v)$, in the enhancement factor integral.

The values of $\langle v_{\textrm{\sc t}}^{2} \rangle$ used in the integral
in equation (\ref{eq3}) were determined from empirical relations
fitted to observational data. From early rocket experiments (e.g.\
Berger et al.\ 1970; Boland et al.\ 1973) to recent results from
\emph{SOHO} (e.g.\ Chae et al.\ 1998; Peter 2001), solar observations
of optically thin TR lines have shown widths in excess of those due to
thermal motion at the temperatures at which the lines are thought to
form. The non-thermal component of the broadening could be due to
unresolved upflows and downflows, to acoustic or MHD waves, or to
small scale turbulence. We interpret the non-thermal widths in the TR
as evidence of turbulent motions of unspecified origin. Since the line
profiles are approximately 
Gaussian, assuming a Gaussian distribution of non-thermal velocities
allows the most probable turbulent velocity $\xi$ to be extracted
from the line width using the expression
\begin{equation}
\frac{\Delta \lambda}{\lambda} = \frac{2\sqrt{\textrm{ln}2}}{c} \left(
\frac{2kT_{\textrm{ion}}}{m_{\textrm{ion}}} + \xi^{2} \right)^{1/2}
\end{equation}
\noindent where $\Delta \lambda$ is the full width at half maximum,
$\lambda$ is the wavelength of the line, $T_{\textrm{ion}}$ is the ion
temperature of line formation, and $m_{\textrm{ion}}$ the mass of the
emitting ion. The mean squared non-thermal velocity $\langle
v_{\textrm{\sc t}}^{2} \rangle$ is given by the relation $\langle
v_{\textrm{\sc t}}^{2} \rangle = (3/2) \xi^{2}$.

Up to log~$T_{\textrm{e}} \sim 5.3$, both recent {\em SOHO} measurements of
$\xi^{2}$ \cite{csl98} and earlier measurements
(see Jordan 1991) are fitted quite well by the relation
$\xi^{2} \propto T_{\textrm{e}}^{1/2}$, which would be expected
if a wave flux from lower regions were conserved through
the transition region to be dissipated at greater heights. The values
of $\langle v_{\textrm{\sc t}}^{2} \rangle (T_{\textrm{i}})$
used in the present study are found using 
\begin{equation}\label{eq17}
\xi = 25 \left( \frac{T_{\textrm{e}}}{2 \times 10^{5} \textrm{ K}}
\right)^{1/4} \textrm{ km s}^{-1}.
\end{equation}
The value of $\xi^{2}$ and hence $\langle v_{\textrm{\sc t}}^{2}
\rangle$ used to characterize the velocity field in each calculation
was taken to be the value given by equation (\ref{eq17}) at the
initial temperature $T_{\textrm{i}}$. 

Peter \shortcite{hp01} has found that some transition region lines
(forming in the temperature range $4.6 \leq$~log~$T_{\textrm{e}} \leq
5.8$) can be fitted more precisely with two Gaussian components. The
widths of the core components are similar to those found by Chae et
al.\ (1998); the widths of the broad components are significantly
larger, but are also consistent with $\xi^{2} \propto
T_{\textrm{e}}^{1/2}$. Peter 
\shortcite{hp01} interprets the broad components as evidence of the
passage of magnetoacoustic waves in coronal funnels in the network. 
It is not known whether the (optically thick) helium lines have this
broad component. If larger turbulent velocities were assumed in the
present calculations, contributions from higher temperatures would
become more important and the enhancement factors derived would be
larger. The temperature dependence of $\xi^{2}$ would mean that the
enhancement factors for He~{\sc ii} would be increased by a relatively
greater amount than those for He~{\sc i}.

\section{Enhancement factor calculations}
\label{sec3}
\subsection{The He~{\sevensize\bf II} 303.8-\AA\ line} 
\label{sec3.1}
Radiative transfer calculations with a 39 level model helium atom
\cite{grs00} show that the contribution function to the intensity
of the 303.8-\AA\ line peaks at log~$T_{\textrm{e}} \simeq 4.9$, and
this value was used for log~$T_{\textrm{i}}$ in the integration of
equation (\ref{eq3}) for the He~{\sc ii} line. The same peak
formation temperature was found by MJ99 using the ionization
equilibrium of Arnaud \& Rothenflug \shortcite{ar85} assuming
purely collisional excitation, which is consistent with the model
calculations. At log $T_{\textrm{i}} = 4.9$, $\sqrt{\langle
v_{\textrm{\sc t}}^{2} \rangle(T_{\textrm{i}})} = 24.3$ km
s$^{-1}$. The collisional excitation rate $C_{ij}(T_{\textrm{e}})$ for
the He~{\sc ii} resonance transition was calculated 
using equation (\ref{eq0}), taking $P_{\textrm{e}}$ as constant and using the
collision strength $\Omega_{ij}(T_{\textrm{e}})$ given by Aggarwal et al.\
\shortcite{aea92}. This rate was also used to calculate the lifetime
of the ground state, $\tau = C_{ij}^{-1}$. The He~{\sc ii} excitation
time at this temperature is 5.69 s, compared with a time of 0.06 s
for the O~{\sc iii} 599.6-\AA\ line, using the collision strengths of
Aggarwal \shortcite{kma93}, and $3 \times 10^{-3}$ s for the C~{\sc
iv} 1548-\AA\ lines \cite{jm92}, formed at slightly higher
temperatures of log~$T_{\textrm{e}} \simeq 5.0$ (these times assume
$P_{\textrm{e}} = 6 \times 10^{14}$ cm$^{-3}$ K). The differences
imply that other lines formed at similar temperatures to the He~{\sc
ii} 303.8-\AA\ line would be relatively unaffected by turbulent transport.

\begin{figure}
\centering
\begin{minipage}{3.6in}
\epsfxsize=3.55in \epsfbox{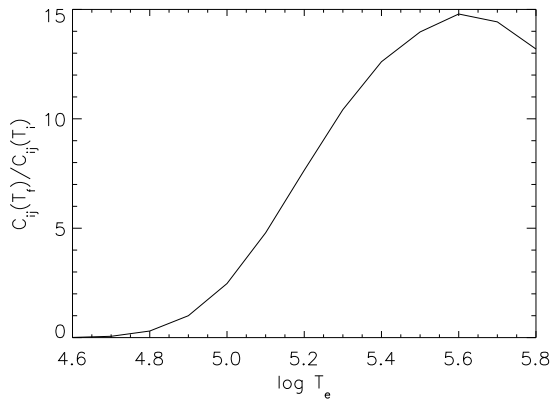}
\end{minipage}
\caption{The excitation rate ratio
$C_{ij}(T_{\textrm{f}})/C_{ij}(T_{\textrm{i}})$ as
a function of (logarithmic) temperature for the He~{\sc ii} 303.8-\AA\
line, as used in calculations of the enhancement
factors. Log~$T_{\textrm{i}} = 4.9$. \label{fig6}}
\end{figure}

\begin{figure}
\centering
\begin{minipage}{3.6in}
\epsfxsize=3.6in \epsfbox{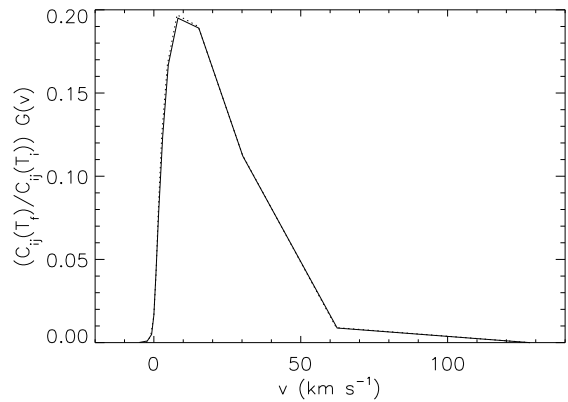}
\end{minipage}
\begin{minipage}{3.6in}
\epsfxsize=3.6in \epsfbox{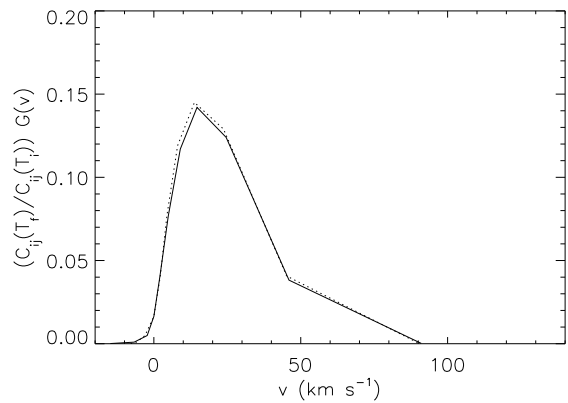}
\end{minipage} 
\caption{The integrand in equation (\protect\ref{eq3}) for the
intensity enhancement, evaluated for the He~{\sc ii}
resonance line, using EMD S (solid) and EMD X (dotted) with
$P_{\textrm{e}} = 2 \times 10^{14}$ cm$^{-3}$ K (top) and
$P_{\textrm{e}} = 6 \times 10^{14}$ cm$^{-3}$ K (bottom). \label{fig2}}
\end{figure}

In order to facilitate a discussion of some of the interesting features
and limitations of the enhancement factor calculations, details of 
the calculations are presented in Table \ref{tab1}
and Figures \ref{fig6} and \ref{fig2}. Table \ref{tab1} shows the
heights above $T_{\textrm{i}}$ corresponding to the 
final temperatures $T_{\textrm{f}}$ determined from the EMDs using equation
(\ref{eq6}). Figure \ref{fig6} shows the excitation rate ratio as a
function of log~$T_{\textrm{e}}$, and Figure \ref{fig2} illustrates
the form of the integrand in equation (\ref{eq3}), showing
the contribution to the total intensity enhancement of plasma elements
with different velocities. The velocities required to reach 
$T_{\textrm{f}}$ are listed in Table
\ref{tab1}. The quantization of log~$T_{\textrm{e}}$ and $h$ produces a fine
mesh of velocities at and below the peak in the excitation rate ratio,
but a coarser grid at higher temperatures. This results in some
over-estimation of the total He~{\sc ii} enhancement factor, 
particularly in the contribution from the material with $v > 30$ km
s$^{-1}$. Rough calculations suggest this over-estimation could be by
about 10--20 per cent.

\begin{table}
\caption{Final temperatures and heights above log~$T_{\textrm{i}}$ = 4.9
reached by plasma elements with velocity $v(T_{\textrm{f}})$, computed using
EMDs S and X. An assumed pressure
of $P_{\textrm{e}} = 2 \times 10^{14}$ cm$^{-3}$ K is denoted by (a),
$P_{\textrm{e}} = 6 \times 10^{14}$ cm$^{-3}$ K is denoted by
(b). Values of $T_{\textrm{f}} < T_{\textrm{i}}$ correspond to
downward-moving elements. \label{tab1}}
\begin{center}
\begin{tabular}{ccccccc}
\hline
log $T_{\textrm{f}}$ & \multicolumn{2}{c}{$h(T_{\textrm{f}}) -
h(T_{\textrm{i}})$} & \multicolumn{4}{c}{$v(T_{\textrm{f}})$} \\
 & \multicolumn{2}{c}{(km)} & \multicolumn{4}{c}{(km s$^{-1}$)} \\ 
 & S & X & S(a) & X(a) & S(b) & X(b) \\\hline
4.6 & -84.4 & -44.8 & -4.94  & -2.62 & -14.8 & -7.87 \\
4.7 & -38.2 & -33.7 & -2.24  & -1.97 & -6.71 & -5.92 \\
4.8 & -13.0 & -17.3 & -0.762 & -1.01 & -2.28 & -3.04 \\
5.0 &  11.6 &  12.3 &  0.680 & 0.721 &  2.04 &  2.16 \\
5.1 &  28.0 &  25.3 &  1.64  & 1.48  &  4.92 &  4.45 \\
5.2 &  51.2 &  45.9 &  3.00  & 2.69  &  9.00 &  8.07 \\
5.3 &  83.9 &  78.6 &  4.92  & 4.60  &  14.7 &  13.8 \\
5.4 &  140  &  135  &  8.20  & 7.91  &  24.6 &  23.7 \\
5.5 &  262  &  257  &  15.3  & 15.1  &  46.0 &  45.2 \\
5.6 &  518  &  513  &  30.3  & 30.1  &  91.0 &  90.2 \\ 
5.7 & 1070  & 1060  &  62.4  & 62.1  &  187  &  186  \\
5.8 & 2190  & 2180  &  128   & 128   &  384  &  383  \\\hline
\end{tabular}
\end{center}
\end{table}

The calculations made assuming $P_{\textrm{e}} = 6 \times 10^{14}$
cm$^{-3}$ K show the peak contribution to the intensity occurring for
upward velocities of about 15 km s$^{-1}$, but with a significant
contribution from faster-moving material. The contribution of material
moving downwards ($v<0$) is seen to be negligible. The peak
contribution to the intensity occurs at log~$T_{\textrm{e}} = 5.3$, at
a height above $T_{\textrm{i}}$ of about 80 km. EMDs S and X lead to
similar temperature profiles, resulting in similar enhancement factors
being derived for each, 5.3 for EMD S and 5.5 for EMD X.

Assuming a lower pressure with these EMDs produces larger enhancement
factors, as predicted in Section \ref{sec2.3}. The total enhancement
factors computed for $P_{\textrm{e}} = 2 \times 10^{14}$ cm$^{-3}$ K
are 6.9 for EMD S and 7.0 for EMD X. Figure \ref{fig2} and Table \ref{tab1}
show that the peak contribution to the intensity occurs at a greater
height and temperature than at the higher pressure, as the velocities
required to reach given $T_{\textrm{f}}$ are decreased (because the
excitation time is longer).

\subsection{The He~{\sevensize\bf I} 584.3-\AA\ and 537.0-\AA\ lines}
Similar calculations were performed for the first two lines of the He~{\sc i}
resonance series, although here the situation is more complicated, as
the coronal approximation no longer applies. First, the He~{\sc i}
resonance lines form at greater optical depths than the He~{\sc ii}
line, and full radiative transfer modelling is required to compute
their intensities properly. Even neglecting transfer of photons within
the line, direct collisional excitation is in general not the dominant
excitation process in radiative transfer calculations \cite{ah69a,aj,grs00}.
Those calculations reveal that important contributions come
from radiative transitions from levels other than the ground state, which
are themselves populated by a combination of collisional excitation
and radiative recombination. Neither the excitation rate nor the mean
lifetime of the He~I ground state can be accurately evaluated using
the equivalent expression to equation (\ref{eq0}) used in the He~{\sc ii}
calculation. In order to investigate the possible effect of
non-thermal motions on the 584.3-\AA\ ($1s2p$~$^{1}P$ -- $1s^{2}$~$^{1}S$)
and 537.0-\AA\ ($1s3p$~$^{1}P$ -- $1s^{2}$~$^{1}S$) lines in a first
approximation, corrections were made to the expressions used to
compute these quantities. Radiative transfer effects were not considered.

Radiative transfer calculations \cite{grs00} reveal that significant
contributions to the 2~$^{1}P$ level population come from allowed
radiative transitions from other singlet levels, especially 2~$^{1}S$,
3~$^{1}S$, and 3~$^{1}D$. These levels are collisionally excited from
the ground at high temperatures but largely populated by radiative
recombination at low temperatures. The collisional contribution to the
population of the  2~$^{1}P$ level through these channels was
approximated in the calculation of the velocity redistribution
enhancement factor by writing the total excitation rate
$C'_{ij}(T_{\textrm{e}})$ as  
\begin{equation}
\label{eq7}
C'_{ij}(T_{\textrm{e}}) = C_{ij}(T_{\textrm{e}}) + \sum_{k \ne j} \left(
\frac{A_{kj}}{\sum_{l}A_{kl}} C_{ik}(T_{\textrm{e}}) \right)
\end{equation}
where $C_{ij}(T_{\textrm{e}})$ is the direct collisional excitation rate, and
the second term is the sum of the collisional excitation rates of the
other contributing levels, multiplied by radiative branching ratios.
$C_{ij}(T_{\textrm{e}})$ and $C_{ik}(T_{\textrm{e}})$ were evaluated
using expression (\ref{eq0}), with collision strengths for
$T_{\textrm{e}} = 3 \times 10^{4}$ K (the collision strengths were
assumed to be constant at higher temperatures, as their variation at
higher $T_{\textrm{e}}$ was often unavailable) provided by Lanzafame et
al.\ \shortcite{lea93} and Sawey \& Berrington \shortcite{sb93}. The
Einstein $A$-values were taken from Drake \shortcite{gd96}.  

The excitation rate ratio was calculated using log~$T_{\textrm{i}} =
4.5$. This temperature was found by MJ99 to represent the peak of
formation of the 584.3-\AA\ line. Radiative transfer calculations
\cite{grs00} suggest a slightly lower temperature log~$T_{\textrm{i}}
\simeq 4.45$ due to a different ionization balance, but as comparisons
are being made here with MJ99, the former figure was kept. Using a
lower $T_{\textrm{i}}$ would have relatively little effect on the
final temperatures responsible for the velocity redistribution
intensity enhancement, as at log~$T_{\textrm{i}} \simeq 4.45$,
excitation times for the He~{\sc i} resonance line are so long (about
40 s; Smith 2000) that ions in the ground state are very likely to
survive at least to log~$T_{\textrm{i}} = 4.5$, and their subsequent
behaviour to be determined by higher temperatures (as for He~{\sc ii},
peak enhancement occurs where log~$T_{\textrm{e}} > 5.0$). The
absolute value of the intensity enhancement is, however, more
significantly affected by the choice of $T_{i}$, through the
excitation rate ratio. The total collisional excitation rate
calculated using equation (\ref{eq7}) is a factor of two smaller at
log~$T_{\textrm{i}} = 4.45$ than at log~$T_{\textrm{i}} = 4.5$, so
choosing a lower $T_{\textrm{i}}$ would be expected to produce a
larger enhancement.

The expression used for the excitation rate is a less exact
approximation to the true rate than in the case of He~{\sc ii},
neglecting as it does the effects of recombination, which is probably
important in the formation of the He~{\sc i} resonance lines
\cite{aj,grs00}. Recombination is most significant below the peak
temperature of line formation. At log~$T_{\textrm{e}} \geq 4.5$, the
temperature range considered here, the expression used represents
about 75 per cent of the total excitation rate computed in radiative
transfer calculations. The fraction does not change significantly
with temperature in this range. As the enhancement factor due to
non-thermal transport of He depends on relative rates at
$T_{\textrm{f}}$ and $T_{\textrm{i}}$, this approximation should have
little effect on the effect on the calculated values of the
enhancement factors. 

Whereas in He~{\sc ii} the collisional excitation of the resonance line
completely dominates the depopulation of the ground state, in He~{\sc i}
there are competing processes which act to shorten the lifetime of the
ground state, some of which do not result in excitation to the
2~$^{1}P$ level. Collisional excitation rates to other bound levels are
significant, as considered above, and collisional ionization becomes
important at high temperatures. Hence the probability that a given
He~{\sc i} ion survives in the ground state until reaching
$T_{\textrm{f}}$ is smaller than would be determined by simply taking
the reciprocal of $C'_{ij}$ (equation (\ref{eq7})). In order to take
this shortening of the \emph{effective} excitation time into account
in a first approximation, for He~{\sc i} $\tau(T_{\textrm{i}})$ was
calculated by removing the radiative branching ratios from the
expression for $C'_{ij}$ and adding the collisional ionization rate
\cite{ms68} before taking the reciprocal to give a mean lifetime (no
longer strictly an `excitation time'). 

\begin{figure}
\centering
\begin{minipage}{3.6in}
\epsfxsize=3.6in \epsfbox{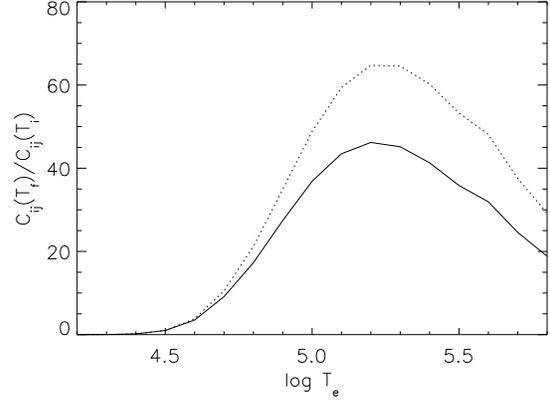}
\end{minipage}
\caption{The excitation rate ratio
$C'_{ij}(T_{\textrm{f}})/C'_{ij}(T_{\textrm{i}})$ as
a function of (logarithmic) temperature for the He~{\sc i} 584.3-\AA\
(solid) and 537.0-\AA\ (dotted) lines, as used in calculations of the
enhancement factors. Log~$T_{\textrm{i}} = 4.5$. \label{fig7}}
\end{figure}

\begin{figure}
\centering
\begin{minipage}{3.6in}
\epsfxsize=3.6in \epsfbox{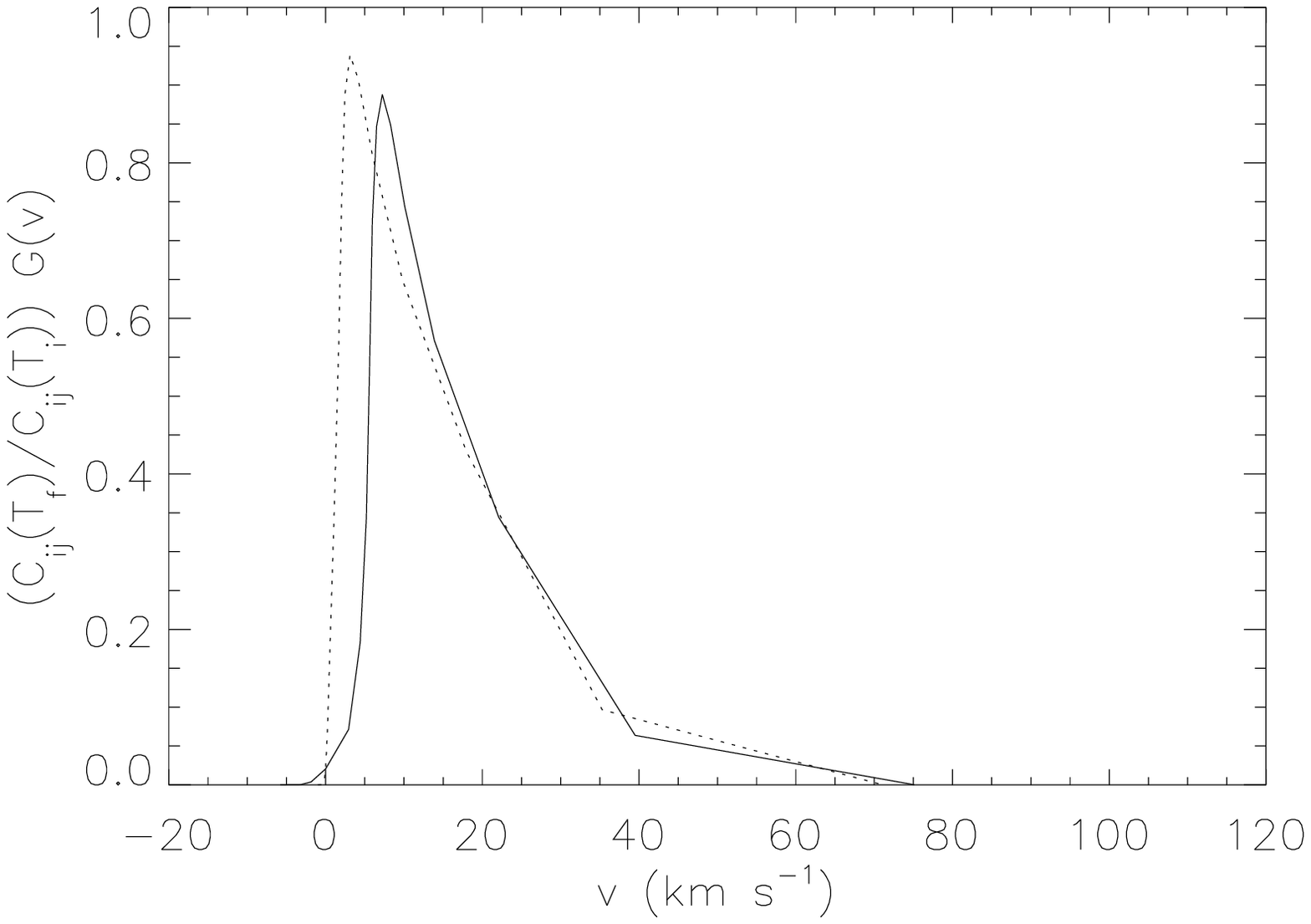}
\end{minipage}
\begin{minipage}{3.6in}
\epsfxsize=3.6in \epsfbox{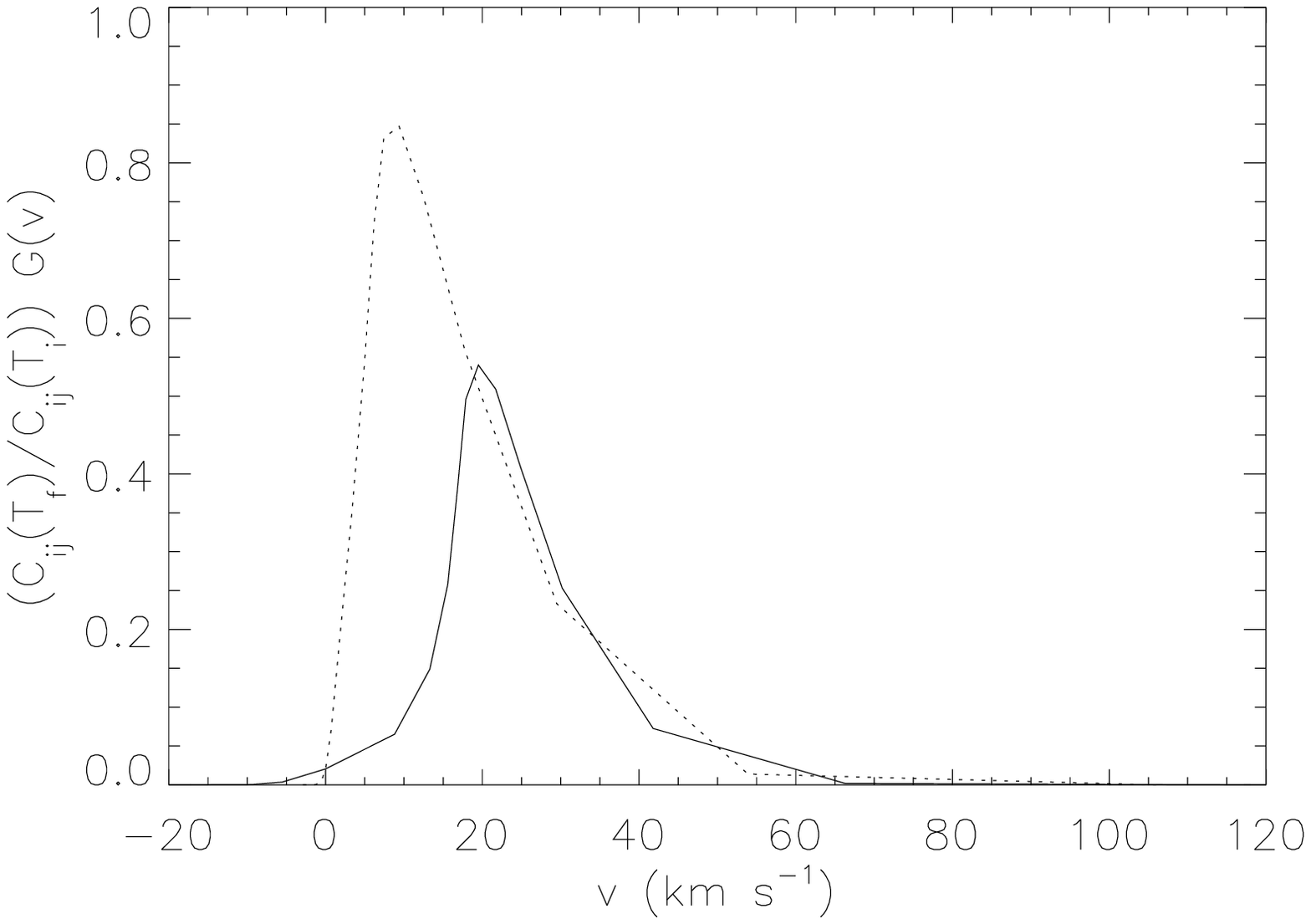}
\end{minipage} 
\caption{The integrand in equation (\protect\ref{eq3}) for the
intensity enhancement, evaluated for the He~{\sc i}
resonance line, using EMD S (solid) and EMD X (dotted) with
$P_{\textrm{e}} = 2 \times 10^{14}$ cm$^{-3}$ K (top) and
$P_{\textrm{e}} = 6 \times 10^{14}$ cm$^{-3}$ K (bottom). \label{fig4}}
\end{figure}

Using these new prescriptions for determining
$C'_{ij}(T_{\textrm{i}})$ and $\tau(T_{\textrm{i}})$, the intensity
enhancement integrals for the 584.3-\AA\ line were performed as for
the He~{\sc ii} resonance line, with $\sqrt{\langle v_{\textrm{\sc
t}}^{2} \rangle(T_{\textrm{i}})} = 19.4$ km s$^{-1}$ and
$\tau(T_{\textrm{i}})$ = 10.5 s (this time may be compared with  
the excitation times, at the same temperature, of the C~{\sc ii} lines
near 1335 \AA\ of about 0.05 s -- Smith 2000). Details
of the integrals are presented in Figures \ref{fig7} and \ref{fig4}
and Table \ref{tab3}. These show that assuming $P_{\textrm{e}} = 6 \times
10^{14}$ cm$^{-3}$ K with EMD S predicts the peak contribution to the
intensity to come from material with $v \simeq 20$ km s$^{-1}$,
reaching log~$T_{\textrm{e}} \simeq 5.1$ at a height of about 200 km above
$T_{\textrm{i}}$. Significant contributions are again seen from
material reaching higher temperatures.
Taking $P_{\textrm{e}} = 2 \times 10^{14}$ cm$^{-3}$ K with EMD S, the
peak contribution comes from slower-moving material ($v < 10$ km
s$^{-1}$), but because the excitation time is longer at the lower
pressure, this material reaches
greater heights and is excited at higher temperatures. The total
enhancement factor is correspondingly increased: 15.3 for
$P_{\textrm{e}} = 2 \times 10^{14}$ cm$^{-3}$ K compared with 11.2 for
$P_{\textrm{e}} = 6 \times 10^{14}$ cm$^{-3}$ K. This can be seen
clearly in Figure \ref{fig4}.

For both values assumed for the pressure, the total enhancement factors
computed using EMD X are larger than for EMD S. A factor of 17.6 is
found for $P_{\textrm{e}} = 2 \times 10^{14}$ cm$^{-3}$ K and a factor
of 18.7 is found for $P_{\textrm{e}} = 6 \times 10^{14}$ cm$^{-3}$
K. In both cases the peak in the intensity contribution occurs at
log~$T_{\textrm{e}} = 5.2$ at a height of about 100 km above
$T_{\textrm{i}}$. The velocities required to reach this
$T_{\textrm{f}}$ are smaller than in the calculations using EMD S, as
the temperature gradients derived from EMD X in the lower TR are
larger. The Gaussian weighting leads to the larger enhancements found
with EMD X. The factors computed for He~{\sc ii} using the two EMDs
are more similar because the (lower TR) EMDs themselves are similar
above the initial temperature (log~$T_{\textrm{i}} = 4.9$). 

\begin{table}
\begin{center}
\caption{Final temperatures and heights above log~$T_{\textrm{i}}$ = 4.5
reached by plasma elements with velocity $v(T_{\textrm{f}})$, computed using
EMDs S and X. An assumed pressure of $P_{\textrm{e}} = 2 \times
10^{14}$ cm$^{-3}$ K is denoted by (a), $P_{\textrm{e}} = 6 \times
10^{14}$ cm$^{-3}$ K is denoted by (b). Values of $T_{\textrm{f}} <
T_{\textrm{i}}$ correspond to downward-moving elements.\label{tab3}}
\begin{tabular}{ccccccc}
\hline
log $T_{\textrm{f}}$ & \multicolumn{2}{c}{$h(T_{\textrm{f}}) -
h(T_{\textrm{i}})$} & \multicolumn{4}{c}{$v(T_{\textrm{f}})$} \\
 & \multicolumn{2}{c}{(km)} & \multicolumn{4}{c}{(km s$^{-1}$)} \\ 
 & S & X & S(a) & X(a) & S(b) & X(b) \\\hline
4.2 & -181.5 & -30.3 & -5.77 & -0.963 & -17.3 & -2.89  \\
4.3 & -99.3  & -13.1 & -3.16 & -0.416 & -9.47 & -1.25  \\
4.4 & -58.1  & -5.95 & -1.85 & -0.189 & -5.54 & -0.567 \\
4.6 &  92.4  &  7.68 &  2.94 &  0.244 &  8.81 &  0.732 \\
4.7 &  139   &  18.8 &  4.42 &  0.597 &  13.3 &  1.79  \\
4.8 &  164   &  35.2 &  5.21 &  1.12  &  15.6 &  3.36  \\
4.9 &  177   &  52.5 &  5.62 &  1.67  &  16.9 &  5.00  \\
5.0 &  188   &  64.8 &  5.97 &  2.06  &  17.9 &  6.18  \\
5.1 &  205   &  77.8 &  6.51 &  2.47  &  19.5 &  7.42  \\
5.2 &  228   &  98.4 &  7.24 &  3.13  &  21.7 &  9.38  \\
5.3 &  261   &  131  &  8.29 &  4.16  &  24.9 &  12.5  \\
5.4 &  317   &  187  &  10.1 &  5.94  &  30.2 &  17.8  \\
5.5 &  439   &  309  &  13.9 &  9.82  &  41.8 &  29.5  \\
5.6 &  695   &  565  &  22.1 &  18.0  &  66.3 &  53.9  \\
5.7 &  1240  &  1110 &  39.5 &  35.3  &  118  &  106   \\
5.8 &  2360  &  2230 &  75.1 &  70.9  &  225  &  213   \\\hline
\end{tabular}
\end{center}
\end{table}

Although the use of log~$T_{\textrm{i}} = 4.5$ for the He~{\sc i}
resonance line may lead to under-estimation of the enhancement factors
if the line forms in equilibrium at a slightly lower temperature,
other approximations used in the calculations tend to lead to
over-estimation. In comparing with the work of MJ99, who assumed line
formation at log~$T_{\textrm{e}} = 4.5$, the figures derived here
should therefore be regarded as upper limits on the possible
enhancement factors. This can be seen in the fact that the enhancement
of the 584.3-\AA\ line is found to increase with pressure in calculations
using EMD X, showing that the uncertainties introduced by the large
velocity steps in the tail of the integrand are significant in this
case. 

MJ99 did not compare the observed intensity of the He~{\sc i}
537.0-\AA\ line directly with the EMD derived
from `normal' transition region lines, but radiative transfer
calculations under-produce the observed intensity of this line by
factors of the same order as in the 584.3-\AA\ line (Smith
2000). Enhancement factors were therefore
computed for the 537.0-\AA\ line, focusing on a comparison with
the 584.3-\AA\ line. Indirect collisional excitation
(mainly through the 2~$^{1}S$, 4~$^{1}S$ and 4~$^{1}D$ levels) was
again included in the total excitation rates calculated for
$T_{\textrm{i}}$ (again log $T_{\textrm{i}} = 4.5$) and
$T_{\textrm{f}}$ and in the calculation of the
mean ground state lifetime. The 537.0-\AA\ line excitation ratio shows
similar variation with velocity (and $T_{\textrm{f}}$) to that of 
the 584.3-\AA\ line, but owing to the higher excitation energy
of the 3~$^{1}P$ level and the intermediate levels important in the
indirect excitation of the 537.0-\AA\ line, the excitation rate ratio
peaks at a slightly higher temperature and falls off more slowly
(relative to its absolute magnitude) than for the 584.3-\AA\ line. 
Very similar calculations
to those performed for the 584.3-\AA\ line were carried out.
The integrated intensity enhancement factors computed for the
537.0-\AA\ line are 14 (EMD S) and 27 (EMD X) for $P_{\textrm{e}} = 6
\times 10^{14}$ cm$^{-3}$ K, and 22 (S) and 26 (X) for $P_{\textrm{e}}
= 2 \times 10^{14}$ cm$^{-3}$ K. The trends exhibited are the same as
those seen in the calculations for the 584.3-\AA\ line. The
ratio of the enhancements computed for the two He~{\sc i} lines is
discussed in Section \ref{sec4}.

\subsection{Non-Maxwellian effects}
\label{sec3.3}
Many studies of the electron velocity distribution function (EVDF) in
the solar transition region have suggested that significant departures
from a Maxwellian distribution may occur (e.g.\ Shoub 1983; Ljepojevic \&
Burgess 1990; Vi\~nas, Wong \& Klimas 2000). This could be due to streaming of
fast electrons from the high TR and corona down the steep temperature
gradient or due to accelerating processes in the chromosphere or
transition region. A form of the EVDF appropriate to the former
process has been used an extensive investigation of the effects of
such a distribution on the helium spectrum (Paper II). Here we
briefly consider the effects of that non-Maxwellian EVDF on the
enhancements predicted by the turbulent transport calculations.

An EVDF that is Maxwellian below a cut off velocity of 3.5
times the thermal velocity, $\sqrt{2kT_{\textrm{e}}/m}$, with a power
law decline with $v^{-31/7}$ \cite{es82} at higher velocities was
considered. This represents a particularly large
departure from the Maxwellian form, and was chosen in order to
produce the maximum effect on collision rates \cite{grs00}. A quantitative
investigation of the effects of such a distribution on collision rates
(and  hence on ground state mean lifetimes) was performed by comparing 
excitation and ionization rates computed in radiative transfer
calculations with and without the non-Maxwellian tail.

This comparison showed that the power law tail in the non-Maxwellian
distribution function does not have a significant effect on the
collision rates at temperatures much above the normal peak temperature
of formation ($T_{\textrm{i}}$). At higher temperatures the excitation
energies of the helium resonance lines correspond to the energies of
electrons in the Maxwellian part of the local distribution, and these
electrons dominate the excitation rates. At and below
$T_{\textrm{i}}$, collisional excitation and ionization occurs
predominantly by electrons in the suprathermal tail of the
distribution, and so rates are significantly enhanced compared with
the Maxwellian case. In the enhancement process investigated here, as
excitation occurs mainly at temperatures much higher than
$T_{\textrm{i}}$, the power law tail has little effect
except at $T_{\textrm{i}}$. The principal effect is to reduce the mean
lifetime of the He~{\sc i} and He~{\sc ii} ground states at their
respective $T_{\textrm{i}}$. For He~{\sc ii} this effect is minimal,
reducing $\tau(T_{\textrm{i}})$ by 10 per cent, but for He~{\sc i} the
reduction is more significant, being by a factor of about 7.  

The increase in $C_{ij}(T_{\textrm{i}})$ in equation (\ref{eq3}) would
also tend to reduce the computed enhancement factor, simply because
emission at $T_{\textrm{i}}$ would be increased; the effect would not
be related to the efficiency of turbulent transport. The change in
$C_{ij}(T_{\textrm{i}})$ is therefore ignored in assessing the effects
of the non-Maxwellian EVDF on the enhancement factor. 

The decrease of $\tau(T_{\textrm{i}})$ has negligible effect on the
enhancements computed for the He~{\sc ii} line, but noticeable effects
on those computed for the He~{\sc i} lines. 
Excitation of the 584.3-\AA\ and 537.0-\AA\ lines occurs
mainly at the same temperatures and heights as in the calculations
with Maxwellian excitation, but the velocities required by plasma
elements to reach those heights are greater. 
There is, therefore, some reduction of the total enhancement
(the same effect occurs for He~{\sc ii}, but to a very much smaller
extent). The reduction is relatively greater in the case of EMD S (the
enhancement factor is smaller by about 30 per cent) than for EMD X, as
the temperature gradient above log~$T_{\textrm{e}} = 4.5$ is smaller
in the former case. The velocity required to reach a particular
temperature is increased for EMD S by a greater amount relative to
$\sqrt{\langle v_{\textrm{\sc t}}^{2} \rangle(T_{\textrm{i}})}$ than
for EMD X, reducing the Gaussian weighting $G(v)$ to a greater extent
in calculating the EMD S enhancement factor. 

In the 537.0-\AA\ line, the effect of the smaller $G(v)$
at high temperatures leads to a reduction in the temperature of the
peak intensity enhancement so that it occurs in the same region as for
the 584.3-\AA\ line. This would tend to reduce any variation in the
ratio of intensities of the two lines compared with the case in which
Maxwellian excitation is assumed. The enhancement of the 537.0-\AA\
line is also decreased, by almost a factor of two in calculations with
EMD S, but by a much smaller amount in the case of EMD X. 

\section{Discussion and conclusions}
\label{sec4}
The intensity enhancement factors calculated for the helium resonance
lines in the presence of turbulent transport are given in Table
\ref{tab5}. As discussed above, these should be regarded as upper limits.
In the case of He~{\sc ii} the over-estimates should be relatively
small, and an intensity enhancement of approximately a factor of 5 is
plausible, given the parameters assumed to describe the solar
atmosphere. This is similar to the factors predicted by Jordan (1980)
and Andretta et al.\ (2000) using a less satisfactory expression for
the excitation time. An enhancement factor of 5 cannot account
entirely for the enhancement of at least a factor of 13 apparently
required in the network according to the analysis of MJ99. However, using
the current second order correction of $sim25$ for the {\sc nis}~2
waveband of {\sc cds} instead of the factor of 55
derived by Landi et al.\ \shortcite{lan97}, the required enhancement
is only a factor of 6 (as found by Jordan 1975), which is within
the range calculated here.
Radiative transfer calculations (Smith 2000) using the VAL
\cite{val} models of the quiet solar atmosphere under-produce the
303.8-\AA\ line intensity by factors of 3--4, which the present work
shows could be explained entirely by the enhancement mechanism
suggested here. 

The enhancement factors derived for the
\hbox{He\,{\sc i}} 584.3-\AA\ line are also upper limits, but show
that enhancement factors of order 10 are plausible. Thus non-thermal
transport of He~{\sc i} could produce the enhancement of the
resonance line intensity required to account for the results of MJ99.

\begin{table}
\caption{Upper limits on intensity enhancement factors for the helium
lines calculated for EMDs S and X. An assumed pressure
of $P_{\textrm{e}} = 2 \times 10^{14}$ cm$^{-3}$ K is denoted by (a),
$P_{\textrm{e}} = 6 \times 10^{14}$ cm$^{-3}$ K is denoted by
(b).\label{tab5}} 
\begin{center}
\begin{tabular}{lccccc}
\hline
Ion & Wavelength & \multicolumn{4}{c}{Enhancement factor}\\
 & (\AA) & S(a) & X(a) & S(b) & X(b) \\\hline
He {\sc ii} & 303.8 & 6.9 & 7.0 & 5.3 & 5.5 \\
He {\sc i}  & 584.3 & 15.3 & 17.6 & 11.2 & 18.7 \\
He {\sc i}  & 537.0 & 22.4 & 26.0 & 14.1 & 26.5 \\\hline
\end{tabular}
\end{center}
\end{table}

Calculations of the 584.3-\AA\ line intensity using the VAL C (average
quiet Sun) and VAL D (average network) models produce fairly good
matches to observations \cite{aj,grs00}, but produce too much emission
in other low transition region lines, owing to the presence of the
temperature plateau at $T_{\textrm{e}} \simeq 2.5 \times 10^{4}$
K. However, models of the atmosphere in which the plateau is removed
do not produce high enough intensities in the \hbox{He\,{\sc i}} line to
match the observations, but by smaller factors than required by
MJ99 \cite{aj,grs00}. These discrepancies could also be explained by
the factors found here.

Similar arguments hold for the \hbox{He\,{\sc i}} 537.0-\AA\ line, for
which an enhancement of a factor of up to about 25 is predicted by the
calculations presented here. It is interesting that greater
enhancement factors are predicted for the 537.0-\AA\ line than for the
584.3-\AA\ line, as radiative transfer calculations using the VAL
models \cite{aj,grs00} show the predicted ratio
$I$(537.0~\AA)/$I$(584.3~\AA) to be smaller than
observed. MJ99 found a ratio of $0.116 \pm 0.015$ in the network
(using the Landi et al.\ 1997 calibration of {\sc cds}; $0.105 \pm
0.014$ using the Brekke et al.\ 2000 calibration), but calculations
using the VAL models give values of about 0.08. Enhancements of the
two lines in the ratio derived here would help resolve this
disagreement. Radiative transfer calculations using a
model atmosphere based on EMD S produce a value of 0.117 (Paper
II); the enhancements predicted here would therefore worsen agreement
with observations. Nevertheless, the enhancement factors calculated
here give an 
intensity ratio much closer to that observed than do radiative
transfer calculations of the effects of non-local suprathermal
electrons which produce enhancement factors of the same order (Paper II).
  
In the formulation used here, the efficiency of the non-thermal
transport process increases with decreasing pressure. For example, the 
predicted 303.8-\AA\ and 584.3-\AA\ line intensity enhancements are
increased by about 30 per cent when the pressure is reduced by a
factor of 3. This would appear to conflict with observations of
coronal holes, where electron pressures are lower than in the quiet
Sun by factors of 2 -- 3 \cite{mw72,jea01}, but where helium line
intensities (and required enhancements) are smaller. However, the
energy balance adopted here in the upper TR (equation (\ref{eq5}))
takes no account of the fast solar wind and the term in $A(r)$ does
not allow for super-radial expansion. For non-thermal transport
calculations appropriate to coronal holes, energy balance calculations
specifically for coronal holes are required; these could be combined
with observed EMDs in a similar manner to that described here.
Further observations
of the helium lines in coronal holes on the disk would be useful (to
avoid the problems of limb darkening/brightening seen in polar coronal
holes). It would also be interesting to investigate non-thermal
transport in active regions (for which Andretta et al.\ 2000 predict
smaller enhancements), where electron pressures are higher
and turbulent velocities lower than in the quiet Sun. 

Turbulent transport could help explain the
spatial variations of the intensities of the helium lines with respect
to each other and to other TR lines in the quiet Sun.
The increases in the heights of peak line formation predicted here
are not large enough to be resolved directly
at the limb, but the effects of the increased
\emph{temperatures} of emission may perhaps be seen in observations of
the network, the appearance of which changes between the low and 
high TR. The turbulent motions of the largely ionized gas would be
influenced by the direction of the network field, so that material
moving to greater heights would trace the expansion of the network
(see e.g.\ Gabriel 1976). 
The width and the contrast of the network observed in the
\hbox{He\,{\sc i}} and \hbox{He\,{\sc ii}} lines seem to resemble that
seen in high TR lines like those of Ne~{\sc vi} (log~$T_{\textrm{e}} =
5.6$) and Ne~{\sc vii} (log~$T_{\textrm{e}} = 5.7$), rather than that
seen in lower TR lines of C~{\sc ii} (log~$T_{\textrm{e}} =
4.4$), C~{\sc iii} (4.9), or even O~{\sc iii} (5.05), O~{\sc iv}
(5.25) or O~{\sc v} (5.4) \cite{bb74,glw76,gea98,pea99}. 
Patsourakos et al.\ (1999) found that Gabriel's (1976) model predicts
a network width which increases with temperature, reaching the width
observed in helium at log~$T_{\textrm{e}} \simeq 5.75$.
The present calculations predict He~{\sc i} emission to peak
at log~$T_{\textrm{e}} \simeq 5.2$, and He~{\sc ii} emission to peak
at log~$T_{\textrm{e}} \simeq 5.3$, but with significant contributions
in each case from temperatures up to log~$T_{\textrm{e}} \simeq 5.6$. 
There is little contribution to intensity above
log~$T_{\textrm{e}} \simeq 5.7$, but the calculations assume
different pressures and temperature gradients (and overall geometry)
to those in the Gabriel (1976) model, making quantitative comparisons
difficult. Qualitatively, however, turbulent transport in an expanding
network could explain why the pattern of helium emission resembles
that seen in lines formed in equilibrium at higher temperatures and
heights. In
a similar geometry, suprathermal electrons streaming down the field
lines from the upper TR would be expected to be funnelled into the
network in the low TR, which would concentrate any enhanced helium
emission in the centre of the network, which would be less consistent
with the observed width of the network.

Turbulent transport of helium in an expanding network is also
qualitatively consistent with observed variations in the ratio of the
intensities of the \hbox{He\,{\sc i}} lines, 
$I$(537.0~\AA)/$I$(584.3~\AA). The excitation rate ratio for the
537.0-\AA\ line peaks at a slightly higher temperature and falls off
more slowly at 
higher temperatures than for the 584.3-\AA\ line. This suggests that
the network could appear wider in the 537.0-\AA\ line than in the
584.3-\AA\ line, which is broadly consistent with rastered images in
the two lines (MJ99), although further observations of the two lines
would be useful. The ratio $I$(537.0~\AA)/$I$(584.3~\AA) would be
expected to increase towards the edges of the network, but the
variation would be relatively small, since the ratio of the
enhancement factors of the two lines changes by only 25 per cent between
log~$T_{\textrm{e}}$ = 4.8 and log~$T_{\textrm{e}}$ = 5.7. This factor
is consistent with values of the ratio observed by MJ99, who found
only small variations of the ratio, but a consistent increase in the
ratio in regions of smaller absolute intensity. Enhanced excitation by
non-local suprathermal electrons in the low TR would tend to produce a
larger variation in the opposite sense with absolute intensities
(Paper II).  

In a more refined approach to turbulent transport calculations,
improvements could be made to the simple geometry adopted
here. Flux tubes with various inclinations could be
modelled, allowing a more quantitative examination of how the
spatial variation of helium emission may be affected. 
MJ99 found steeper temperature gradients in cell interiors than in the
network, but these gradients may be \emph{across} magnetic
structures. In the expanding network picture, the temperature gradient
\emph{along the field} is smaller, which would reduce the enhancement
factor, as would the higher pressures found by MJ99 in cell interiors. 
The greater helium line enhancements required by MJ99's
observations in cell interiors could possibly be caused by
photon scattering from the boundaries; radiative transfer using
two-component model atmospheres is needed to test this point. 

As described in Section \ref{sec3.3}, even significantly
non-Maxwellian EVDFs are expected to have relatively
little effect on turbulent transport of helium in the quiet Sun. As
enhancements of the helium line intensities by non-thermal motions
occur at temperatures much higher than the normal temperature of peak
line formation, and enhancement by non-local suprathermal electrons
occurs largely at lower temperatures (Paper II), the two processes
could be operating simultaneously. If non-thermal motions
do produce the enhancement factors calculated here, these would
dominate over the effects of non-local electrons, provided
departures from Maxwellian EVDFs are realistic (Paper II). 

The present calculations do not include radiative transfer, which
could have significant effects on \hbox{He\,{\sc i}}, but which would
be less important for \hbox{He\,{\sc ii}}. Radiative
transfer calculations including non-thermal motions (and possibly
non-Maxwellian EVDFs as well) are needed to place our conclusions on a
firmer footing.

\section*{Acknowledgments}
GRS acknowledges the financial support of PPARC as a DPhil student,
under grant PPA/S/S/1997/02515.

\end{document}